\documentclass[12pt]{article}

\input epsf.tex
\newcommand{\be}{\begin{equation}}
\newcommand{\ee}{\end{equation}}
\newcommand{\bea}{\begin{eqnarray}}
\newcommand{\eea}{\end{eqnarray}}

\begin{document}

\bigskip 
\begin{titlepage}

\begin{flushright}
UUITP-23/03\\ 
hep-th/0312203
\end{flushright}

\vspace{1cm}

\begin{center}
{\Large\bf A matrix model black hole: act II\\}

\end{center}
\vspace{3mm}

\begin{center}

{\large
Ulf H.\ Danielsson} \\

\vspace{5mm}

Institutionen f\"or Teoretisk Fysik, Box 803, SE-751 08
Uppsala, Sweden

\vspace{5mm}

{\tt
ulf@teorfys.uu.se \\
}

\end{center}
\vspace{5mm}

\begin{center}
{\large \bf Abstract}
\end{center}
\noindent
In this paper we discuss the
connection between the deformed matrix model and two dimensional black holes in the light
of the new developements involving fermionic type 0A-string theory. We argue that many of the old
results can be carried over to this new setting and that the original claims about the deformed
matrix model are essentially correct. We show the agreement between correlation functions calculated using 
continuum and matrix model techniques. We also explain how detailed properties of the space
time metric of the extremal blck hole of type 0A are reflected in the deformed matrix model.


\vfill
\begin{flushleft}
December 2003
\end{flushleft}
\end{titlepage}
\newpage


\section{Introduction}

\bigskip

Recently, there has been a renewed interest in the $c=1$ matrix model in the
context of fermionic type 0 strings in two dimensions, see, e.g., [1-7].
Originally, the matrix model was constructed in order to describe strings as
continuum limits of triangulated world sheets, for a nice review see \cite
{Klebanov:1991qa}. The new developments show how the matrix model emerges as
the theory for open string tachyons living on a collection of D-particles in
the two dimensional space time. With this in mind, it is interesting to go
back to some of the unanswered questions left behind by the first matrix
revolution.

One such question concerns a possible matrix model for a two dimensional
black hole that was proposed in \cite{Jevicki:1993zg}: the deformed matrix
model. The model includes a $1/\lambda ^{2}$ contribution to the matrix
potential and is, like the usual $c=1$ matrix model, exactly solvable. After
its introduction the model was studied in a number of papers, see ,e.g.,
[10-15], and successfully passed several important tests. Interestingly, the
deformed matrix model has now turned up again as the matrix model for
fermionic type 0A strings in the presence of D-flux, \cite{Douglas:2003up}.
In this paper we will review some of the old results on the deformed matrix
model and explain their connection with the 0A model. Our conclusion will be
that the original interpretation of the deformed matrix model proposed in 
\cite{Jevicki:1993zg}  is basically correct.

The outline of the paper is as follows. In section two we discuss the basics
of string theory in flat two dimensional space time. For the convenience of
the reader we first review some of the results from the first matrix
revolution -- both from a continuum and a matrix model point of view. We
also briefly review how the 0A/0B models fit into the story. In section
three we turn to the black hole. First we discuss the original claims that
the deformed matrix model described a bosonic black hole and the evidence in
favor. We then turn to the extremal black hole of 0A to see how the
arguments carry over to this much better controlled case. We end with some
conclusions.

\bigskip

\section{Flat space time}

\bigskip

\subsection{The bosonic c=1 theory}

\bigskip

\subsubsection{In the continuum}

We begin, for the benefit of the reader, with a brief review of the bosonic
string in two dimensions. A nice introduction to the subject with references
and many original results, can be found in \cite{DiFrancesco:1991ud}. The
space time action for the metric, the dilaton $\Phi $ and the tachyon $T$,
is given by 
\begin{equation}
S\sim \int d^{2}x\sqrt{-G}\left[ e^{-2\Phi }\left( \frac{16}{\alpha ^{\prime
}}+R+4\left( \nabla \Phi \right) ^{2}-\frac{1}{2}\left( \nabla T\right) ^{2}+%
\frac{2}{\alpha ^{\prime }}T^{2}\right) \right] ,
\end{equation}
up to interaction terms. Varying the action we obtain apart from the
Einstein equations for the metric, a dilaton equation of motion given by 
\begin{equation}
\frac{16}{\alpha ^{\prime }}+R-4\left( \nabla \Phi \right) ^{2}+4\nabla
^{2}\Phi =0,
\end{equation}
and a tachyon equation of motion given by 
\begin{equation}
\nabla ^{2}T-2\nabla \Phi \cdot \nabla T+\frac{4}{\alpha ^{\prime }}T=0.
\end{equation}
Flat space time is a solution if we allow for a linear dilaton background
according to 
\begin{equation}
\Phi =-\frac{2}{\sqrt{\alpha ^{\prime }}}\phi ,  \label{lindil}
\end{equation}
which implies a string coupling that increases towards negative $\phi $.
This non-trivial dilaton background is the prize we have to pay for not
having $26$ dimensions.

From the point of view of the world sheet theory, the linear dilaton
corresponds to a background charge which shifts the counting of the central
charge allowing for two space time dimensions instead of $26$. Another way
to arrive at the same theory, is to start with just one scalar field, the $%
c=1$ theory, and investigate the physics of the two dimensional world sheet
metric induced by the quantum mechanical breaking of the world sheet
conformal invariance. The metric has only one degree of freedom, the
Liouville mode, corresponding to a Weyl-scaling of the metric. This is what
we above interpreted as an extra space dimension making the original one
dimensional $c=1$ model effectively two dimensional.

The next step is to study the propagation of tachyons in this background.
Plugging the linear dilaton background, and a flat metric, into the tachyon
equation of motions with the ansatz $T=e^{ikX+\beta \phi }$, one finds 
\begin{equation}
\frac{\alpha ^{\prime }}{4}k^{2}-\frac{\alpha ^{\prime }}{4}\beta ^{2}-\sqrt{%
\alpha ^{\prime }}\beta =1,  \label{masskal}
\end{equation}
with solutions 
\begin{equation}
\beta =-\frac{2}{\sqrt{\alpha ^{\prime }}}\pm \left| k\right| ,  \label{beta}
\end{equation}
if $X$ is Euclidean time. With $X$ as ordinary time we find 
\begin{equation}
\beta =-\frac{2}{\sqrt{\alpha ^{\prime }}}\pm ik,
\end{equation}
i.e. left and right movers in space time. In the following we will be
working in Euclidean time, and we will focus on tachyons where we choose the 
$+$-sign in (\ref{beta}) -- i.e. what in the matrix literature is referred
to as the `right' gravitational dressing. We note that after a rescaling by $%
e^{-\frac{2}{\sqrt{\alpha ^{\prime }}}\phi }$ the tachyons correspond to
massless excitations. (In two dimensions the name tachyon is, therefore, a
bit of a misnomer).

We now proceed to consider a background with a tachyon condensate given by 
\begin{equation}
T_{0}=\Delta e^{-\frac{2}{\sqrt{\alpha ^{\prime }}}\phi },  \label{nolltach}
\end{equation}
where $\Delta $, which controls the magnitude of the tachyon condensate, can
be identified with the world sheet cosmological constant. The condensate
increases towards negative $\phi $ and effectively gives rise to a wall --
the tachyon wall -- against which the tachyons can scatter. The idea is to
send in tachyons from the weakly coupled regime at $\phi \rightarrow +\infty 
$ and see what happens. Of particular interest are correlation functions for
tachyons where all tachyons, except one, has the same sign of the momentum
(all other vanishes, \cite{DiFrancesco:1991ud}). We then have, at genus
zero, 
\begin{equation}
\left\langle T_{k_{1}}...T_{k_{N-1}}T_{-k_{N}}\right\rangle \sim
\prod_{n=1}^{N}\gamma \left( 1-\sqrt{\alpha ^{\prime }}k_{n}\right) \frac{%
d^{N-3}}{d\mu ^{N-3}}\mu ^{\sum_{n=1}^{N}\sqrt{\alpha ^{\prime }}k_{n}/2-1},
\label{boscorr}
\end{equation}
where $\gamma \left( x\right) \equiv \frac{\Gamma \left( x\right) }{\Gamma
\left( 1-x\right) }$, and instead of $\Delta $, we have written the
correlation function as a function of $\mu $. The relation between $\Delta $
and $\mu $ will be explained below.

There are several comments that have to be made. Usually the pole structure
of string amplitudes is very involved, but in two dimensions the kinematics
leads to substantial simplifications. As we see, the poles come in the form
of completely factorized leg factors. This has to do with the fact that it
is only when tachyons of opposite chirality come together that an on shell
intermediate state is created. Second, the calculation of the correlation
function can only be performed perturbatively in $\Delta $. That is, the
calculation of a correlation function at non zero $\Delta $ reduces to a
calculation at zero $\Delta $ with an extra number of zero momentum
tachyons, \cite{Goulian:1990qr}. As a consequence, we can only perform the
calculation when the power of $\Delta $ is expected to be an integer. For
other values we must rely on analytical continuation. Each such tachyon is,
just as all the other tachyons in (\ref{boscorr}), accompanied with a leg
factor which in this case is $\gamma \left( 1\right) =0$. Hence, the
correlation function naively vanishes. This is interpreted as the need for a
renormalization of $\Delta $ and is the reason for why the introduction of a
new parameter $\mu $ is convenient. As argued in, e.g. \cite{Klebanov:1991qa}
and \cite{DiFrancesco:1991ud}, the zeroes of the leg poles can be regulated
to $\ln \mu $, with $\Delta =\mu \ln \mu $, leading to (\ref{boscorr}) as
promised.

As emphasized above, the continuum calculation only works for 
\begin{equation}
\sum_{n=1}^{N}\sqrt{\alpha ^{\prime }}k_{n}/2-1=s,
\end{equation}
with $s$ integer, and it follows, therefore, from the conservation laws 
\begin{eqnarray}
\sum_{n=1}^{N}\left( -\frac{2}{\sqrt{\alpha ^{\prime }}}+\left| k_{n}\right|
\right) -\frac{2s}{\sqrt{\alpha ^{\prime }}} &=&-\frac{4}{\sqrt{\alpha
^{\prime }}} \\
\sum_{n=1}^{N-1}k_{n} &=&k_{N},
\end{eqnarray}
that 
\begin{equation}
k_{N}=\frac{N-2+s}{\sqrt{\alpha ^{\prime }}}.  \label{momon}
\end{equation}
This means that we will always sit on one of the poles of the leg-factor
associated with the $N$'th tachyon. The pole can be regulated as $%
\lim_{\varepsilon \rightarrow 0}\frac{\mu ^{s+\varepsilon }}{\varepsilon }=%
\frac{\mu ^{s}}{0}+\mu ^{s}\ln \mu $, where the first term can be thought of
as a UV divergence that can be renormalized away from the world sheet point
of view, or, alternatively, as an infinite volume term from the space time
point of view. As explained above, we are, in space time, working on a semi
infinite line cut off towards negative $\phi $ by the tachyon wall. The
position of the wall, i.e. the point where $T_{0}$ becomes of order one, is
given by $\phi \sim \frac{\sqrt{\alpha ^{\prime }}}{2}\ln \mu $.

Due to the way that the correlation function factorizes, it is convenient to
absorb the leg factors into the normalization of the tachyons and write 
\begin{equation}
T_{k}=\widetilde{T}_{k}\gamma \left( 1-\sqrt{\alpha ^{\prime }}k_{n}\right) .
\end{equation}
The only part of the correlation function that remains, is now the
non-factorizable piece. For instance, if we consider the case $s=0$ we
simply have 
\begin{equation}
\left\langle \widetilde{T}_{k_{1}}...\widetilde{T}_{k_{N-1}}\widetilde{T}%
_{-k_{N}}\right\rangle =\left( N-3\right) !.
\end{equation}

Let us finally come back to the mass-shell equation (\ref{masskal}). From
the world sheet point of view this equation has a very simple
interpretation: it is nothing but the condition that the world sheet vertex
operators should have conformal dimension equal to $\left( 1,1\right) $.
Generalizing to operators that include world sheet derivatives, one finds
the condition 
\begin{equation}
\frac{\alpha ^{\prime }}{4}k^{2}-\frac{\alpha ^{\prime }}{4}\beta ^{2}-\sqrt{%
\alpha ^{\prime }}\beta =1-n,  \label{masskaln}
\end{equation}
where we have $n$ world sheet derivatives $\partial $, and the same number
of derivatives $\overline{\partial }$. In two dimensions there are no
transverse degrees of freedom and one would naively not expect any physical
states to arise in this way. However for some special, discrete values of $k$%
, the gauge transformations collapse, [18-23,28]. This happens at $k=\frac{n%
}{\sqrt{\alpha ^{\prime }}}$ for $n$ integer. It is precisely these special
values of momenta that show up in correlation functions, as is apparent from
the leg factors above. One such example is $k=0$ at $n=1$, where the two
solutions of (\ref{masskaln}) are given by 
\begin{eqnarray}
V_{1,1}^{(+)} &=&\partial X\overline{\partial }X  \label{gravdress} \\
V_{1,1}^{(-)} &=&\partial X\overline{\partial }Xe^{-\frac{4}{\sqrt{\alpha
^{\prime }}}\phi },  \nonumber
\end{eqnarray}
(up to gauge equivalent changes of polarization) and correspond to two
different gravitational dressings of the operators, the right and wrong
dressings respectively. The first corresponds, in this particular example,
to changing the radius of compactification, the second, as we will see, is
related to a space time black hole.

\subsubsection{The matrix model}

We now turn to the matrix model description of the $c=1$ two dimensional
string. As argued in [24-27], the appropriate matrix model has the potential 
\begin{equation}
V\left( x\right) =-\frac{1}{2\alpha ^{\prime }}\lambda ^{2},  \label{bosmat}
\end{equation}
where the Fermi sea is filled up to $\mu $ from the top of the inverted
harmonic oscillator. Sometimes in the literature, one uses a Fermi sea that
is filled only on one side of the potential, but for non-perturbative
stability it is convenient to fill both sides. A useful, and very intuitive
trick, is to calculate scattering amplitudes using analytic continuation $%
\alpha ^{\prime }\rightarrow -\alpha ^{\prime }$ from the right side up
harmonic potential. This leads, for instance, to the two puncture (i.e. zero
energy tachyon) correlation function, 
\begin{equation}
\left\langle PP\right\rangle =-\frac{1}{\pi }\rm{Im}\sum_{n=0}\frac{1}{%
E_{n}-\mu },
\end{equation}
where 
\begin{equation}
E_{n}=\frac{i}{2\sqrt{\alpha ^{\prime }}}\left( 2n+1\right) .
\end{equation}
The advantage of this way of doing things is that the poles in the tachyon
correlation function can be seen to have a very simple explanation, \cite
{Gross:1990js}\cite{Danielsson:1991bc}\cite{Danielsson:1991jf}. Simple
perturbation theory shows that the poles correspond to energy differences, $%
E_{n}-E_{m},$ in the harmonic oscillator which indeed give rise to poles at $%
\frac{n}{\sqrt{\alpha ^{\prime }}}$ for $n$ integer. Calculations in the
matrix model furthermore precisely reproduce the results of (\ref{boscorr})
for the correlation functions. The non-factorizable piece is identically the
same, and the pole structure of the legs also agrees.

\bigskip

\subsection{The 0A/0B theories}

\bigskip

As explained in the introduction, it was recently realized that many of the
results for the bosonic string have an analogue in the case of
non-supersymmetric fermionic strings, i.e. the 0A and 0B models. In the case
of the 0B-string, there are two space time scalars, a massless tachyon and a
massless RR-scalar. They both have an on-shell condition that can be
obtained from (\ref{masskal}) simply by replacing $\alpha ^{\prime
}\rightarrow 2\alpha ^{\prime }$, i.e. 
\[
\frac{\alpha ^{\prime }}{4}k^{2}-\frac{\alpha ^{\prime }}{4}\beta ^{2}-\sqrt{%
\frac{\alpha ^{\prime }}{2}}\beta =\frac{1}{2}, 
\]
with solutions 
\[
\beta =-\sqrt{\frac{2}{\alpha ^{\prime }}}\pm \left| k\right| . 
\]
Of particular importance is the fact that, just as in the bosonic case, we
have a matrix model description of the theory. The 0B matrix model is
immediately obtained from the bosonic model by replacing $\alpha ^{\prime
}\rightarrow 2\alpha ^{\prime }$ in (\ref{bosmat}). The new twist of the
story, as explained in e.g. \cite{Douglas:2003up}, is that we now have a
much better understanding of from where the matrix model comes. In the 0B
theory there are unstable D-particles in the space time. The sign of the
instability, is the presence of open string tachyons with end points living
on the D-particles. The matrices describing these tachyons are nothing else
than the matrices of the matrix model, and the upside down potential is a
signal of the tachyonic behavior.

The presence in the 0B theory of two types of scalars have an interesting
explanation from the point of view of the matrix model potential. As
explained in \cite{Douglas:2003up}, perturbations with even parity in the
Fermi sea correspond to tachyons, while perturbations with odd parity
correspond to the RR-scalars. Continuum calculations furthermore show that
the tachyon correlation functions have poles at $k=\frac{2n}{\sqrt{2\alpha
^{\prime }}}$ while the RR-scalars have poles at $k=\frac{2n+1}{\sqrt{%
2\alpha ^{\prime }}}$. Since the energy levels of the harmonic oscillator
now are given by 
\begin{equation}
E_{n}=\frac{i}{2\sqrt{2\alpha ^{\prime }}}\left( 2n+1\right) ,
\end{equation}
these poles correspond to transitions between states of equal and different
parity, respectively, just as one would expect.

In the case of 0A strings the situation is slightly different. We no longer
have any RR-scalar and, therefore, only the tachyon remains. As shown in 
\cite{DiFrancesco:1991ud} the genus zero tachyon correlation functions now
take the form, for $s=0$,

\begin{equation}
\left\langle T_{k_{1}}...T_{k_{N-1}}T_{-k_{N}}\right\rangle =\left(
N-3\right) !\prod_{n=1}^{N}\gamma \left( 1-\sqrt{\frac{\alpha ^{\prime }}{2}}%
k_{n}\right) .  \label{0acorr}
\end{equation}
We first note that the non-factorizable piece is the same as in the bosonic
case, furthermore we note that only half of the poles are present\ in the
leg factors compared with the bosonic case. This has an important
consequence. Momentum conservation force the last tachyon to have $k_{N}=%
\frac{N-2}{\sqrt{2\alpha ^{\prime }}}$ (compare with (\ref{momon})) which
leads to a leg factor $\gamma \left( 2-\frac{N}{2}\right) $ which diverges
only for $\mathit{even}$ $N$. We conclude, therefore, that only correlation
functions with an even number of tachyons are non-vanishing. We will come
back to the matrix model description of the 0A string in the next section,
after a review of the two dimensional black hole.

\section{The black hole}

\bigskip

\subsection{The bosonic black hole}

\bigskip

In \cite{Jevicki:1993zg} it was conjectured that the bosonic two dimensional
black hole, \cite{Witten:1991yr}, corresponds to a matrix model with the
potential

\begin{equation}
V\left( \lambda \right) =-\frac{1}{2\alpha ^{\prime }}\lambda ^{2}+\frac{M}{%
2\lambda ^{2}}.
\end{equation}
There were several promising indications in this direction, of which the
most obvious are the scaling properties of the theory. If we reinsert
Planck's constant into the matrix model Schr\"{o}dinger equation, we find,
near the top of the Fermi sea,

\begin{equation}
\left( -\frac{\hbar ^{2}}{2}\frac{d^{2}}{d\lambda ^{2}}-\frac{1}{2\alpha
^{\prime }}\lambda ^{2}+\frac{M}{2\lambda ^{2}}\right) \psi \left( \lambda
\right) =\mu \psi \left( \lambda \right) .
\end{equation}
We now rescale $\lambda $, by writing $\lambda =x\sqrt{\hbar }$, to get all
terms to be of same order in $\hbar $. We then find 
\begin{equation}
\left( -\frac{1}{2}\frac{d^{2}}{dx^{2}}-\frac{1}{2\alpha ^{\prime }}x^{2}+%
\frac{M\hbar ^{-1}}{2x^{2}}\right) \psi \left( x\right) =\mu \hbar
^{-1/2}\psi \left( x\right) .
\end{equation}
This suggests that the string coupling scales like $1/\mu $, or $1/\sqrt{M}$%
, which is just what we would expect from comparing (\ref{nolltach}) and (%
\ref{gravdress}) with (\ref{lindil}).

Another important point is the position of the poles. As explained in \cite
{Danielsson:1993wq}, all solutions of the Schr\"{o}dinger equation are not
admissable. When $M>3/4$ the even parity wave functions blow up near the
origin, and only the odd parity ones should be kept. This is the case which
will be relevant for us. \ In passing we note that for -$1/4<M<3/4$ there is
a one parameter family of self-adjoint extensions that can be used, while in
the case of $M<-1/4$, finally, the energies turn complex signalling an
instability. In our case, with a positive $M$, one is naturally lead to the
throwing away of all the even parity states and, as a consequence, we loose
half of the poles. Interestingly, this is precisely what happens if one
scatters tachyons in the black hole background, compared with the scattering
of tachyons in a background with a tachyon condensate, as shown in \cite
{Jevicki:1993zg}.

As already mentioned, all calculations that can be made in the ordinary
matrix model, can also be done in the deformed matrix model. Including
tachyon correlation functions to all genus. Matrix model calculations of
correlation functions lead to the following result at genus zero, see, e.g., 
\cite{Danielsson:1994ac}, where the leg factors have been factorized out: 
\begin{eqnarray}
\left\langle \widetilde{T}_{k_{1}}...\widetilde{T}_{k_{N-1}}\widetilde{T}%
_{-k_{N}}\right\rangle  &=&\left( N-3\right) !!\left( \sqrt{\alpha ^{\prime }%
}k_{N}-2\right) \left( \sqrt{\alpha ^{\prime }}k_{N}-4\right) ...  \nonumber
\\
&&...\left( \sqrt{\alpha ^{\prime }}k_{N}-\left( N-4\right) \right) M^{k_{N}/%
\sqrt{\alpha ^{\prime }}-N/2+1}.  \label{deftach}
\end{eqnarray}

An important question is whether these results can be verified using
continuum methods. In \cite{Bershadsky:1991in} it was argued that the
bosonic black hole can be reproduced by adding the operator $V_{1,1}^{(-)}$,
i.e. a \textit{black hole screener}, to the world sheet action for a string
in flat space time. In fact, by expanding the black hole metric around a
flat background one recovers the black hole screener. In \cite
{Bershadsky:1991in}, however, it was argued that the addition of the black
hole screener to the action was an \textit{exact} description of the black
hole of \cite{Witten:1991yr}. This was the approach used in \cite
{Danielsson:1994sk} to calculate correlation functions perturbatively in $M$%
. The calculations are in general technically difficult, but there are some
cases where the calculations can be done with less effort. In \cite
{Danielsson:1994sk} values for momenta near $k_{N}=\frac{N}{\sqrt{\alpha
^{\prime }}}$, $k_{1}=...=k_{N-1}=\frac{1}{\sqrt{2\alpha ^{\prime }}}$, were
chosen in order to perform the calculation with one black hole screener. The
advantage of this choice is that one can arrange so that $\zeta \cdot
k_{i}=0 $ for $i\leq N-2$, where $\zeta $ is the polarization of the black
hole screener. In this way the matrix model result above, in the case $k_{N}/%
\sqrt{\alpha ^{\prime }}-N/2+1=1$, was reproduced, giving rise to 
\begin{equation}
\left\langle \widetilde{T}_{k_{1}}...\widetilde{T}_{k_{N-1}}\widetilde{T}%
_{-k_{N}}\right\rangle =\left( N-2\right) !M\gamma \left( 1\right) ,
\end{equation}
where 
\begin{equation}
T_{k}=\gamma \left( 1-\sqrt{\alpha ^{\prime }}k\right) \widetilde{T}_{k}.
\end{equation}
Hence we find complete agreement for the non-factorizable part with (\ref
{deftach}).

There are a couple of important observations that one can should make.
First, there are poles for \textit{all} discrete momenta in this continuum
calculation. In \cite{Danielsson:1994sk} it was suggested that this was due
to the choice of vacuum. In the matrix model it is simple to project out the
odd parity states. In the continuum a similar procedure is needed which,
actually is nothing else than what happens when one go from 0B to 0A.

Second, there is a zero provided by the formal factor $\gamma \left(
1\right) $. This is just the same phenomena that is encountered for the
tachyon background. Again we need a rescaling -- like the one that brings us
from $\Delta $ to $\mu $ -- in order to obtain a non-zero correlation
function. As we will see, this can be understood in a very nice way from the
form of the space time metric.

\subsection{The extremal black hole}

\bigskip

\subsubsection{The continuum picture}

\bigskip

Let us now turn to the case of the 0A-string. We begin by recalling the
space time action for the 0A-theory with a D-flux $q$: 
\[
S\sim \int d^{2}x\sqrt{-G}\left[ e^{-2\Phi }\left( \frac{8}{\alpha ^{\prime }%
}+R+4\left( \nabla \Phi \right) ^{2}-\frac{1}{2}\left( \nabla T\right) ^{2}+%
\frac{1}{\alpha ^{\prime }}T^{2}\right) -\frac{q^{2}}{4\pi \alpha ^{\prime }}%
\right] . 
\]
Systems of this form has been studied several times in the literature, see,
e.g., \cite{McGuigan:1991qp}\cite{Berkovits:2001tg}. Following \cite
{Berkovits:2001tg} we solve for the dilaton and the metric (without tachyon
condensate) obtaining: 
\begin{eqnarray*}
ds^{2} &=&g\left( \phi \right) dt^{2}+\frac{1}{g\left( \phi \right) }d\phi
^{2} \\
\Phi &=&-\sqrt{\frac{2}{\alpha ^{\prime }}}\phi \\
g\left( \phi \right) &=&1-e^{-2\sqrt{\frac{2}{\alpha ^{\prime }}}\phi
}\left( m\sqrt{\frac{\alpha ^{\prime }}{2}}+\frac{q^{2}}{16\pi }\sqrt{\frac{2%
}{\alpha ^{\prime }}}\phi \right) .
\end{eqnarray*}
Note that the dilaton is still linear as in (\ref{lindil}) thanks to a
convenient choice of gauge. We will be particularly interested in the
extremal case, which we expect will correspond to a stable ground state of
the system. The metric above has two horizons corresponding to the two
solutions of $g\left( \phi \right) =0$. The two horizons coincide when $%
g^{\prime }\left( \phi \right) =0$, with the double horizon situated at 
\[
\phi _{0}=-\frac{1}{2}\sqrt{\frac{\alpha ^{\prime }}{2}}\ln \frac{32\pi }{%
q^{2}}. 
\]
As a consequence, the string coupling on the horizon is given by 
\begin{equation}
g_{s}=\frac{\sqrt{32\pi }}{q},
\end{equation}
and decreases exponentially as we move away. The mass parameter, finally, is
given by 
\begin{equation}
m_{0}=\frac{q^{2}}{32\pi }\sqrt{\frac{2}{\alpha ^{\prime }}}\ln \frac{32\pi e%
}{q^{2}}.  \label{mq}
\end{equation}
We will come back to this important relation a little bit later -- it is of
crucial importance when we want to identify the deformed matrix model with
the extremal black hole.

One concludes from the above that by choosing a large value for $q$ the
closed string coupling can be made arbitrarily small everywhere outside of
the horizon. On the other hand, if we want to make calculations involving
open strings attached to the $q$ D-particles, the effective coupling is $%
qg_{s}$ which is fixed and not small. In \cite{Berkovits:2001tg} it was
noted that the near horizon limit of the extremal black hole is an $AdS_{2}$
space. The finite value of $qg_{s}$ will, presumably, have important
consequences for any analogue in two dimensions of the AdS/CFT
correspondence in higher dimensions.\footnote{%
While this work was being completed, a similar remark was made by A.
Strominger in a talk at the Santa Barbara workshop on Strings and Cosmology.}

\bigskip

\subsubsection{The matrix model}

\bigskip

The 0A-background described in the previous section can, according to \cite
{Douglas:2003up}, be described using a matrix model with the potential 
\begin{equation}
V\left( x\right) =-\frac{1}{4\alpha ^{\prime }}\lambda ^{2}+\frac{q^{2}-%
\frac{1}{4}}{2\lambda ^{2}}.
\end{equation}
Again one can argue for this choice of potential from the point of view of
open string tachyons, \cite{Douglas:2003up}. In 0A D-particles are stable,
contrary to the case of 0B, but, luckily, it is instead possible to consider
systems of D-particles and anti D-particles. The tachyons stretching between
the particles and anti particles become complex and the Fermi sea two
dimensional. The angular coordinate can, however, be integrated out and the
resulting potential is the one given above, with $q$ as the net number of
D-particles. The difference in the coefficient in front of the $\lambda ^{2}$
can, as in the case of 0B, be understood from the fact that the open string
tachyon has $m^{2}=-\frac{1}{\alpha ^{\prime }}$ in case of the bosonic
string and $m^{2}=-\frac{1}{2\alpha ^{\prime }}$ in case of the 0A.

Following \cite{Danielsson:1993wq} we may formally continue $\alpha ^{\prime
}\rightarrow -\alpha ^{\prime }$ in order to have a right side up harmonic
oscillator. We then have the following Schr\"{o}dinger equation to solve: 
\begin{equation}
\left( -\frac{1}{2}\frac{d^{2}}{d\lambda ^{2}}+\frac{1}{4\alpha ^{\prime }}%
\lambda ^{2}+\frac{\eta }{2\lambda ^{2}}\right) \psi \left( x\right) =E\psi
\left( x\right) ,
\end{equation}
where $\eta =q^{2}-1/4$. The energy eigenvalues can be shown to be 
\begin{equation}
E_{n}=\frac{i}{2\sqrt{2\alpha ^{\prime }}}\left( 2n+1+2a_{n}\right) ,
\end{equation}
where 
\begin{equation}
a_{n}=-(-1)^{n}\left( -\frac{1}{2}+\sqrt{\frac{1}{4}+\eta }\right) .
\end{equation}
The even energy levels, with $n=2l$, can now be written 
\begin{equation}
E_{2l}=\frac{1}{2\sqrt{2\alpha ^{\prime }}}\left( 4l+1+1-q\right) =\frac{1}{%
\sqrt{2\alpha ^{\prime }}}\left( 2l+1-\frac{q}{2}\right) ,
\end{equation}
and the odd ones, with $n=2l+1$, 
\begin{equation}
E_{2l+1}=\frac{1}{2\sqrt{2\alpha ^{\prime }}}\left( 4l+3-1+q\right) =\frac{1%
}{\sqrt{2\alpha ^{\prime }}}\left( 2l+1+\frac{q}{2}\right) .
\end{equation}
As we have already explained, for $q\geq 0$, we must throw away the even
parity solutions leaving only the odd levels\footnote{%
Note that for $q^{2}<0$, i.e. an imaginary D-flux, the solutions are
unstable. This case was studied in \cite{Gross:2003zz} in the context of
T-duality between the 0A and the 0B-model.}. \ Repeating the argument of 
\cite{Danielsson:1991jf}\cite{Danielsson:1992bx} we predict poles at $\frac{%
2n}{\sqrt{2\alpha ^{\prime }}}$, in agreement with the continuum results
above.

It is also gratifying to see that the somewhat mysterious critical value for
the black hole mass of $-1/4$, that we observed in the previous section, now
is shifted to zero. This, by the way, throws light on the properties of the
deformed matrix model under T-duality. The ordinary $c=1$ matrix model at
finite radius was first studied in \cite{Gross:1990ub}, while the deformed
model has been studied in \cite{Demeterfi:1993sj} and \cite
{Danielsson:1993dh}. A convenient way to proceed with the analysis is to
consider the puncture two point function at finite radius written as 
\begin{equation}
\left\langle PP\right\rangle =\frac{1}{\pi R}\rm{Im}\sum_{n,m=0}\frac{1}{%
\left( E_{n}+\frac{2m+1}{R}-\mu \right) ^{2}}.
\end{equation}
In this way T-duality is obvious at each genus (recall that the genus
expansion corresponds to an expansion in $1/\mu ^{2}$) and involves an
exchange of $n$ and $m$ in the sum. If we now, as in \cite{Danielsson:1993dh}%
, do the same thing for the deformed model (remembering that we only should
keep odd states) we run into a problem if we consider $1/\eta $ above as the
genus counting parameter. As observed in \cite{Demeterfi:1993sj} T-duality
seems, in this case, to be broken. In \cite{Danielsson:1993dh} a modified
transformation was proposed, which, as observed in \cite{Kapustin:2003hi},
corresponds to regarding $1/q^{2}$ as the parameter that keeps track of the
genus.

\subsubsection{Comparing the deformed matrix model with the extremal black
hole}

\bigskip

It would be interesting to repeat and extend the calculations made in \cite
{Danielsson:1994sk}, and discussed in section 3.1, to the fermionic type 0A
string. We will not do this here, but instead take a different approach to
reproduce the deformed matrix model results. Given the factorized
expressions, both for the bosonic case and the case of the 0A, it is
actually quite easy to derive the non factorizable piece of the scattering
amplitude with an arbitrary number of black hole screeners inserted without
the evaluation of any complicated integrals. In the following we use
conventions appropriate for the 0A string but the argument is identical in
the case of the bosonic string. We start with 
\[
\left\langle \widetilde{T}_{k_{1}}...\widetilde{T}_{k_{N-1}}\widetilde{T}%
_{-k_{N}}\right\rangle \sim \mu ^{\sum_{n=1}^{N}\sqrt{\frac{\alpha }{2}%
^{\prime }}k_{n}+2-N}.
\]
and take a $\mu $-derivative to insert a puncture, $P$, that is, a zero
momentum tachyon. We find 
\[
\left\langle P\widetilde{T}_{k_{1}}...\widetilde{T}_{k_{N-1}}\widetilde{T}%
_{-k_{N}}\right\rangle \sim \left( \sum_{n=1}^{N}\sqrt{\frac{\alpha ^{\prime
}}{2}}k_{n}+2-N\right) \frac{1}{\mu }\left\langle \widetilde{T}_{k_{1}}...%
\widetilde{T}_{k_{N-1}}\widetilde{T}_{-k_{N}}\right\rangle .
\]
From this we identify terms including $\sqrt{\alpha ^{\prime }}k_{i}/2$
corresponding to the puncture being attached to the $i$'th tachyon leg,
while the rest of the pre-factor comes from the puncture being attached
somewhere else in the diagram. What if we want to insert a black hole
screener $V_{1,1}^{(-)}$ instead? From momentum conservation, we note that
two zero momentum tachyons can fuse together into one black hole screener.
Hence, if we have a diagram with a puncture already present we should attach
another one to the same leg. In this way we effectively insert a black hole
screener in the rest of the diagram. \ Using the result above we expect that
the insertion of a black hole screener can be accomplished by the limit 
\[
\left\langle \widetilde{T}_{k_{1}}...\widetilde{T}_{k_{N-1}}\widetilde{T}%
_{-k_{N}}V_{1,1}^{(-)}\right\rangle \sim \lim_{k_{1}\rightarrow 0}\frac{k_{1}%
}{\mu }\left\langle P\widetilde{T}_{k_{1}}...\widetilde{T}_{k_{N-1}}%
\widetilde{T}_{-k_{N}}\right\rangle +...
\]
The zero that we encounter is the same $\gamma \left( 1\right) $ that we
already has seen in the case of the insertion of zero momentum tachyons, and
is something that we need to compensate for. It is, \cite{DiFrancesco:1991ud}%
, a general result that the insertion of wrongly dressed special operators
make the correlation functions vanish.

So, we have found that a black hole screener can be inserted by acting with $%
\frac{1}{\mu }\frac{d}{d\mu }$ on the correlation function. Based on this we
now propose that 
\begin{eqnarray}
\left\langle \widetilde{T}_{k_{1}}...\widetilde{T}_{k_{N-1}}\widetilde{T}%
_{-k_{N}}V_{1,1}^{(-)m}\right\rangle  &=&\left\langle \widetilde{T}_{k_{N}}%
\widetilde{T}_{-k_{N}}P^{N-2}V_{1,1}^{(-)m}\right\rangle   \nonumber \\
&\sim &\left( \frac{d}{d\mu }\right) ^{N-2}\left( \frac{1}{\mu }\frac{d}{%
d\mu }\right) ^{m}\left\langle \widetilde{T}_{k_{N}}\widetilde{T}%
_{-k_{N}}\right\rangle \gamma \left( 1\right) ^{m},  \label{svartkorr}
\end{eqnarray}
where the $\gamma \left( 1\right) $ must be aborbed into the coefficient of
the $V_{1,1}^{(-)}$ in the world sheet action, just as in the case of the
tachyon condensate. However, there is still one thing that remains to be
explained. There is in principle a question of ordering between the
operation of inserting a puncture and a black hole screener using our
recipe. Since we have that 
\[
\left[ \frac{d}{d\mu },\frac{1}{\mu }\frac{d}{d\mu }\right] =-\frac{1}{\mu
^{2}}\frac{d}{d\mu },
\]
the ordering actually do matter. Luckily, however, the difference is easy to
interpret. It simply corresponds to a diagram where \textit{three} punctures
come together on the same leg. That is, a puncture is attached to an
external graviton leg. If we reverse the order between the derivatives in (%
\ref{svartkorr}) we will miss such contributions and the result will,
therefore, not be correct. \ When we now are confident that (\ref{svartkorr}%
) is correct, we evaluate it to find 
\[
\left( \frac{d}{d\mu }\right) ^{N-2}\left( \frac{1}{\mu }\frac{d}{d\mu }%
\right) ^{m}\frac{\mu ^{N-2+2m}}{N-2+2m}=\left( N-3\right) !!\left(
N+2m-4\right) !!
\]
in agreement with the results derived in, e.g., \cite{Danielsson:1994sk}.
Another convenient way to summarize the results is to write, as in \cite
{Danielsson:1994ac}, 
\[
\left\langle \widetilde{T}_{k}\widetilde{T}_{-k}\right\rangle _{M,\mu }\sim 
\frac{\left( M+\mu ^{2}\right) ^{\sqrt{2\alpha ^{\prime }}k}}{k},
\]
where $M=q^{2}$.\footnote{%
It is amusing to note that the $k\rightarrow 0$ limit of this expression,
i.e. the two puncture correlation funtion, can be written as $\ln \left(
M+\mu ^{2}\right) =\ln \left( \mu +iq\right) +\ln \left( \mu -iq\right) $.
This is precisely the expressions used in \cite{Gross:2003zz} on the 0B side
after T-duality, if one puts $q=iQ$.}

The results above are independent of whether we are discussing the bosonic
string or the 0A. In the bosonic case, the field theory calculation in \cite
{Danielsson:1994sk} did not include a projection to remove all the odd
states, implying, incorrectly, that correlation functions with an odd number
of tachyons are non-zero. In the 0A however, and presumably also for a
correctly projected bosonic string, the leg factors are of the form given in
(\ref{0acorr}). As a consequence, the leg factor corresponding to tachyon
number $N$, which we have chosen to have the opposite chirality compared
with the others, is given by $\gamma \left( 2-N/2\right) $ which diverges
only for \textit{even} $N$.

Finally, we need to fulfill our promise and come back to the relation
between the mass and the charge for the extremal black hole derived in (\ref
{mq}). Up to the logarithm we note that the mass of the black hole goes like
the square of the charge. That is, the coefficient of the $1/\lambda ^{2}$
is indeed (up to the shift of $1/4$) the black hole mass, just as
conjectured! Furthermore, we should note that the expression $m\sim q^{2}\ln
q^{2}$ is a perfect analogue to the expression $\Delta \sim \mu \ln \mu $
that we have discussed earlier. $\Delta $ and $m$ are parameters natural for
the string world sheet physics, while $\mu $ and $q$ appear naturally in the
matrix model. In this way we see that the formal $\gamma \left( 1\right) =0$
is compensated for by a volume factor just as in case of the tachyon
condensate.

\section{Conclusions}

\bigskip

In this paper we have discussed the space time interpretation of the
deformed matrix model. We have found that the old conjecture that the model
describes a two dimensional black hole comes to new life in the context of
the 0A string. There are many things left to do. For instance, there are
more special states at discrete momenta than the one corresponding to a
black hole space time. In the matrix model one can, in principle, calculate
correlation functions also in the presence of those. But what is the space
time interpretation from the point of view of the 0A?

\section*{Acknowledgments}

The author would like to thank Martin Olsson and Peter Rajan for
discussions. The author is a Royal Swedish Academy of Sciences Research
Fellow supported by a grant from the Knut and Alice Wallenberg Foundation.
The work was also supported by the Swedish Research Council (VR).


\bigskip

\begin{thebibliography}{99}
\bibitem{takayanagi}  T.~Takayanagi and N.~Toumbas, ``A matrix model dual of
type 0B string theory in two dimensions,'' JHEP \textbf{0307}, 064 (2003)
[arXiv:hep-th/0307083].

\bibitem{McGreevy:2003kb}  J.~McGreevy and H.~Verlinde, ``Strings from
tachyons: The c = 1 matrix reloated,'' arXiv:hep-th/0304224.

\bibitem{Martinec:2003ka}  E.~J.~Martinec, ``The annular report on
non-critical string theory,'' arXiv:hep-th/0305148.

\bibitem{Klebanov:2003km}  I.~R.~Klebanov, J.~Maldacena and N.~Seiberg,
``D-brane decay in two-dimensional string theory,'' JHEP \textbf{0307}, 045
(2003) [arXiv:hep-th/0305159].

\bibitem{McGreevy:2003ep}  J.~McGreevy, J.~Teschner and H.~Verlinde,
``Classical and quantum D-branes in 2D string theory,'' arXiv:hep-th/0305194.

\bibitem{Douglas:2003up}  M.~R.~Douglas, I.~R.~Klebanov, D.~Kutasov,
J.~Maldacena, E.~Martinec and N.~Seiberg, ``A new hat for the c = 1 matrix
model,'' arXiv:hep-th/0307195.

\bibitem{DeWolfe:2003qf}  O.~DeWolfe, R.~Roiban, M.~Spradlin, A.~Volovich
and J.~Walcher, ``On the S-matrix of type 0 string theory,'' JHEP \textbf{%
0311} (2003) 012 [arXiv:hep-th/0309148].

\bibitem{Klebanov:1991qa}  I.~R.~Klebanov, ``String theory in
two-dimensions,'' arXiv:hep-th/9108019.

\bibitem{Jevicki:1993zg}  A.~Jevicki and T.~Yoneya, ``A Deformed matrix
model and the black hole background in two-dimensional string theory,''
Nucl.\ Phys.\ B \textbf{411}, 64 (1994) [arXiv:hep-th/9305109].

\bibitem{Danielsson:1993wq}  U.~H.~Danielsson, ``A Matrix model black
hole,'' Nucl.\ Phys.\ B \textbf{410} (1993) 395 [arXiv:hep-th/9306063].

\bibitem{Demeterfi:1993sj}  K.~Demeterfi and J.~P.~Rodrigues, ``States and
quantum effects in the collective field theory of a deformed matrix model,''
Nucl.\ Phys.\ B \textbf{415}, 3 (1994) [arXiv:hep-th/9306141].

\bibitem{Demeterfi:1993cm}  K.~Demeterfi, I.~R.~Klebanov and
J.~P.~Rodrigues, ``The Exact S matrix of the deformed c = 1 matrix model,''
Phys.\ Rev.\ Lett.\ \textbf{71}, 3409 (1993) [arXiv:hep-th/9308036].

\bibitem{Danielsson:1993dh}  U.~H.~Danielsson, ``The deformed matrix model
at finite radius and a new duality symmetry,'' Phys.\ Lett.\ B \textbf{325}
(1994) 33 [arXiv:hep-th/9309157].

\bibitem{Danielsson:1994ac}  U.~H.~Danielsson, ``Two-dimensional string
theory, topological field theories and the deformed matrix model,'' Nucl.\
Phys.\ B \textbf{425} (1994) 261 [arXiv:hep-th/9401135].

\bibitem{Danielsson:1994sk}  U.~H.~Danielsson, ``The Scattering of strings
in a black hole background,'' Phys.\ Lett.\ B \textbf{338} (1994) 158
[arXiv:hep-th/9405052].

\bibitem{DiFrancesco:1991ud}  P.~Di Francesco and D.~Kutasov, ``World sheet
and space-time physics in two-dimensional (Super)string theory,'' Nucl.\
Phys.\ B \textbf{375}, 119 (1992) [arXiv:hep-th/9109005].

\bibitem{Goulian:1990qr}  M.~Goulian and M.~Li, ``Correlation Functions In
Liouville Theory,'' Phys.\ Rev.\ Lett.\ \textbf{66} (1991) 2051.

\bibitem{Gross:1990js}  D.~J.~Gross, I.~R.~Klebanov and M.~J.~Newman, ``The
Two Point Correlation Function Of The One-Dimensional Matrix Model,'' Nucl.\
Phys.\ B \textbf{350} (1991) 621.

\bibitem{Polyakov:qx}  A.~M.~Polyakov, ``Selftuning Fields And Resonant
Correlations In 2-D Gravity,'' Mod.\ Phys.\ Lett.\ A \textbf{6} (1991) 635.

\bibitem{Danielsson:1991bc}  U.~H.~Danielsson and D.~J.~Gross, ``On the
correlation functions of the special operators in c = 1 quantum gravity,''
Nucl.\ Phys.\ B \textbf{366} (1991) 3.

\bibitem{Klebanov:1991hx}  I.~R.~Klebanov and A.~M.~Polyakov, ``Interaction
of discrete states in two-dimensional string theory,'' Mod.\ Phys.\ Lett.\ A 
\textbf{6}, 3273 (1991) [arXiv:hep-th/9109032].

\bibitem{Danielsson:1991jf}  U.~H.~Danielsson, ``Symmetries and special
states in two-dimensional string theory,'' Nucl.\ Phys.\ B \textbf{380}
(1992) 83 [arXiv:hep-th/9112061].

\bibitem{Polyakov:1991xa}  A.~M.~Polyakov, ``Singular states in 2-d quantum
gravity,'' \textit{Lectures given at the 1991 Jerusalem Winter School}.

\bibitem{Gross:1990ay}  D.~J.~Gross and N.~Miljkovic, ``A Nonperturbative
Solution Of D = 1 String Theory,'' Phys.\ Lett.\ B \textbf{238}, 217 (1990).

\bibitem{Brezin:1989ss}  E.~Brezin, V.~A.~Kazakov and A.~B.~Zamolodchikov,
``Scaling Violation In A Field Theory Of Closed Strings In One Physical
Dimension,'' Nucl.\ Phys.\ B \textbf{338}, 673 (1990).

\bibitem{Ginsparg:1990as}  P.~Ginsparg and J.~Zinn-Justin, ``2-D Gravity +
1-D Matter,'' Phys.\ Lett.\ B \textbf{240}, 333 (1990).

\bibitem{Parisi:1990jy}  G.~Parisi, ``String Theory On The One-Dimensional
Lattice,'' Phys.\ Lett.\ B \textbf{238} (1990) 213.

\bibitem{Danielsson:1992bx}  U.~H.~Danielsson, ``A Study of Two-Dimensional
String Theory,'' Princeton PhD. Thesis, arXiv:hep-th/9205063.

\bibitem{Witten:1991yr}  E.~Witten, ``On string theory and black holes,''
Phys.\ Rev.\ D \textbf{44} (1991) 314.

\bibitem{Bershadsky:1991in}  M.~Bershadsky and D.~Kutasov, ``Comment on
gauged WZW theory,'' Phys.\ Lett.\ B \textbf{266} (1991) 345.

\bibitem{McGuigan:1991qp}  M.~D.~McGuigan, C.~R.~Nappi and S.~A.~Yost,
``Charged black holes in two-dimensional string theory,'' Nucl.\ Phys.\ B 
\textbf{375} (1992) 421 [arXiv:hep-th/9111038].

\bibitem{Berkovits:2001tg}  N.~Berkovits, S.~Gukov and B.~C.~Vallilo,
``Superstrings in 2D backgrounds with R-R flux and new extremal black
holes,'' Nucl.\ Phys.\ B \textbf{614}, 195 (2001) [arXiv:hep-th/0107140].

\bibitem{Gross:2003zz}  D.~J.~Gross and J.~Walcher, ``Non-perturbative RR
potentials in the c(hat) = 1 matrix model,'' arXiv:hep-th/0312021.

\bibitem{Gross:1990ub}  D.~J.~Gross and I.~R.~Klebanov, ``One-Dimensional
String Theory On A Circle,'' Nucl.\ Phys.\ B \textbf{344}, 475 (1990).

\bibitem{Kapustin:2003hi}  A.~Kapustin, ``Noncritical superstrings in a
Ramond-Ramond background,'' arXiv:hep-th/0308119.
\end{thebibliography}
\end{document}